\documentclass[aps,prb,superscriptaddress,showpacs,floatfix,twocolumn]{revtex4}

\usepackage{graphicx,graphics}
\usepackage{dcolumn}
\usepackage{amsmath,amssymb,amsfonts}
\usepackage{latexsym,verbatim}
\usepackage{bm}
\usepackage{color}
\usepackage{ulem}

\def\be{\begin{equation}}
\def\ee{\end{equation}}
\def\ber{\begin{eqnarray}}
\def\eer{\end{eqnarray}}

\begin{document}
\title{Double-layer graphene and topological insulator thin-film plasmons}
\author{Rosario E.V. Profumo}
\affiliation{Dipartimento di Fisica dell'Universit\`a di Pisa and NEST, Istituto Nanoscienze-CNR, I-56127 Pisa, Italy}
\author{Reza Asgari}
\email{asgari@ipm.ir}
\affiliation{School of Physics, Institute for Research in Fundamental Sciences (IPM), Tehran 19395-5531, Iran}
\author{Marco Polini}
\email{m.polini@sns.it}
\homepage{http://qti.sns.it}
\affiliation{NEST, Istituto Nanoscienze-CNR and Scuola Normale Superiore, I-56126 Pisa, Italy}
\author{A.H. MacDonald}
\affiliation{Department of Physics, University of Texas at Austin, Austin, Texas 78712, USA}
\begin{abstract}
We present numerical and analytical results for the optical and acoustic plasmon
collective modes of coupled massless-Dirac two-dimensional electron systems. 
Our results apply to topological insulator (TI)  thin films and to two graphene sheets separated by a 
thin dielectric barrier layer.  We find that because of strong bulk dielectric screening TI acoustic modes are locked to the top of the 
particle-hole continuum and therefore probably unobservable. 
\end{abstract}


\maketitle

\section{Introduction}
\label{sect:intro}

The physics of closely-spaced but unhybridized two-dimensional electron systems (2DESs)
has been a subject of theoretical and experimental interest since it was first
appreciated~\cite{pogrebinskii_1977,price_physicaB_1983} that electron-electron interactions
allow energy and momentum to be transferred between layers, while maintaining separate particle-number conservation.
Remote Coulomb coupling has commanded a great deal of attention 
during the past thirty years or so because it provides a potential alternative to the 
inductive and capacitive coupling of conventional electronics.
Until recently, remote Coulomb coupling research focused on quasi-2D electron systems 
confined to nearby quantum wells in molecular-beam-epitaxy grown semiconductor heterostructures.
The study of Coulomb-coupled 2D systems has now been revitalized
by advances which have made it possible to prepare robust and ambipolar 2DESs,
based on graphene~\cite{graphenereviews} layers
or on the surface states of topological insulators~\cite{TIreviews},
that are described by an ultrarelativistic wave equation instead of the non-relativistic Schr\"{o}dinger equation.  

Single- and few-layer graphene systems can be produced by mechanical exfoliation
of thin graphite or by thermal decomposition of silicon carbide~\cite{SiCreviews}. 
Isolated graphene layers host massless-Dirac two-dimensional electron systems 
(MD2DESs) with a four-fold  (spin $\times$ valley)  flavor degeneracy,
whereas topologically-protected MD2DESs that have no additional spin or valley flavor
labels appear automatically~\cite{TIreviews,zhang_natphys_2009} at the top and bottom surfaces
of a three-dimensional (3D) TI thin film.  
The protected surface states of 3D TIs are associated with spin-orbit interaction driven bulk band inversions.  3D TIs in a slab geometry offer two surface states that can be far enough
apart to make single-electron tunneling negligible, but close enough for Coulomb interactions between surfaces to be important.
Unhybridized MD2DES pairs can be realized in graphene by separating two 
layers by a dielectric~\cite{kim_prb_2011} (such as ${\rm Al}_2{\rm O}_3$) 
or by a few layers of a one-atom-thick insulator such as BN~\cite{dean_naturenano_2010,ponomarenko_naturephys_2011}. 
In both cases inter-layer hybridization is negligible and the nearby graphene layers are, from the point of view of single-particle physics, isolated.  Isolated graphene layers can be also found on the surface of bulk graphite~\cite{grapheneongraphite,li_natphys_2009} and in ``folded graphene"~\cite{schmidt_prb_2010}  (a natural byproduct of micromechanical exfoliation), or prepared by chemical vapor deposition~\cite{li_natphys_2009}. 
We use the term {\it double-layer graphene} (DLG) to
refer to a system with two graphene layers that are coupled only by Coulomb 
interactions, avoiding the term {\it bilayer graphene} which 
typically refers to two adjacent graphene layers in the crystalline Bernal-stacking configuration~\cite{borghi_prb_2009}.

DLG and TI thin films are both described at low energies by a Hamiltonian 
with two MD2DES~\cite{graphenereviews} coupled only by Coulomb interactions. 
The importance of electron-electron interactions in MD2DESs  
has been becoming more obvious as sample quality has improved~\cite{eeinteractionsgraphene}, 
motivating investigations of charge and spin or pseudospin dynamics in DLG and thin-film TIs in 
the regime in which long-range Coulomb forces give rise to robust plasmon collective
modes~\cite{Pines_and_Nozieres,Giuliani_and_Vignale}. 
Because of their electrically tunable collective behaviors, 
DLG and thin-film TIs may have a large impact on  {\it plasmonics}, 
a very active subfield of optoelectronics~\cite{Ebbesen_PT_2008,Maier07,koppens_nanolett_2011}
whose aim is to exploit plasmon properties in order to compress infrared
electromagnetic waves to the nanometer scale of modern electronic devices.

In this Article we use the random phase approximation (RPA)~\cite{Pines_and_Nozieres,Giuliani_and_Vignale} 
to evaluate the optical and acoustic plasmon mode dispersions in DLG and in thin-film TIs. 
In particular, we obtain an {\it exact} analytical formula for the RPA acoustic plasmon group velocity
valid for arbitrary substrate and barrier dielectrics that points to a key difference between these two MD2DES's, namely that the velocity in 
TI thin films is strongly suppressed.  The RPA collective modes of DLG have been calculated earlier by Hwang and Das Sarma~\cite{hwang_prb_2009}: below we will comment at length on the relation between our results and theirs. Plasmon collective modes formed from TI surface states have also been considered 
previously by Raghu {\it et al.}~\cite{raghu_prl_2010} in the regime in which coupling between top and bottom surfaces can
be neglected. Based on our analysis, we are able to clarify how dielectric screening influences plasma 
frequencies in this limit.  

Plasmons can be observed by a variety of experimental tools including inelastic light scattering~\cite{pellegrini_review_2006}, which has been widely used to probe plasmons in semiconductor heterostructures~\cite{plasmonsILSsemiconductors}, but also by surface-physics techniques like high-resolution electron-energy-loss spectroscopy~\cite{liu_prb_2008}, and, more indirectly, angle-resolved photoemission spectroscopy~\cite{eeinteractionsgraphene}. Double-layer field-effect transistors with a grating gate~\cite{peralta_apl_2002} can also be used to detect plasmons. Coupling between far-infrared light and Dirac plasmons in single-layer graphene has recently been achieved by employing an array of graphene nanoribbons~\cite{ju_naturenano_2011} and by performing near-field scanning optical microscopy through the tip of an AFM~\cite{fei_nanolett_2011}.

This manuscript is organized as follows. In Sect.~\ref{sect:model} we present the model we have used to describe a pair of 
Coulomb-coupled MD2DESs, and introduce the linear-response functions which describe collective electron dynamics. 
In Sect.~\ref{sect:collectivemodes} we present and discuss our main analytical and numerical results for the dispersion of optical and acoustic plasmons in these systems. Finally, in Sect.~\ref{sect:conclusions} we present 
a summary of our main conclusions. 

\begin{figure}
\centering
\includegraphics[width=0.40\linewidth]{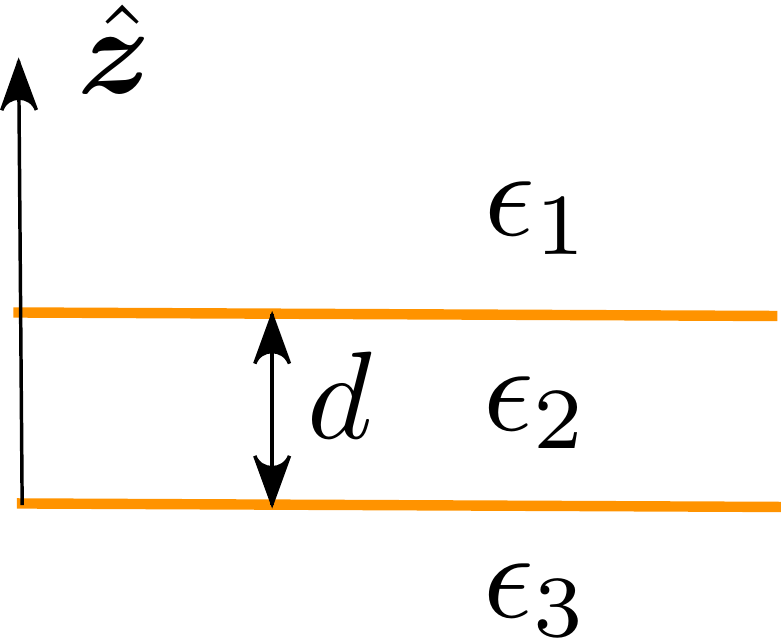}
\caption{(Color online) A side view of the double-layer system described by Eq.~(\ref{eq:Hamiltonian}), 
which explicitly indicates the dielectric model used in these calculations. The two layers hosting massless Dirac fermions are located at $z=0$ and $z=d$.\label{fig:one}}
\end{figure}

%
\section{Model Hamiltonian and Random Phase Approximation}
\label{sect:model}

We consider two unhybridized MD2DESs separated by a finite distance $d$ and embedded in the dielectric
environment depicted in Fig.~\ref{fig:one}. 
The two systems are assumed to be coupled solely by Coulomb interactions. 
The Hamiltonian describing this system reads~\cite{profumo_prb_2010} ($\hbar =1$)
\ber\label{eq:Hamiltonian}
{\hat {\cal H}} &=&  v \sum_{{\bm k}, \ell, \alpha, \beta} {\hat \psi}^\dagger_{{\bm k}, \ell, \alpha} 
( {\bm \sigma}_{\alpha\beta} \cdot {\bm k} ) {\hat \psi}_{{\bm k}, \ell, \beta} \nonumber\\
&+& \frac{1}{2 S}\sum_{{\bm q}, \ell, \ell'} V_{\ell \ell'}(q){\hat \rho}_{{\bm q}, \ell} {\hat \rho}_{-{\bm q}, \ell'}~.
\eer
Here $v$ is the bare Dirac velocity, taken to be the same in the $\ell = 1,2$ tunnel-decoupled layers, 
$S$ is the area of each layer, $V_{\ell \ell'}(q)$ is the matrix of bare Coulomb potentials, and 
\begin{equation}\label{eq:densityoperator}
{\hat \rho}_{{\bm q}, \ell} = \sum_{{\bm k}, \alpha} {\hat \psi}^\dagger_{{\bm k} - {\bm q}, \ell, \alpha}{\hat \psi}_{{\bm k}, \ell, \alpha}
\end{equation}
is the density-operator for the $\ell$-th layer. 
The Greek letters are honeycomb-sublattice-pseudospin labels and ${\bm \sigma} = (\sigma^x,\sigma^y)$ is a vector of Pauli matrices. 
A sum over flavor labels is implicit in Eq.~(\ref{eq:densityoperator}) in the case of DLG. 
The relative strength of Coulomb interactions is measured by the dimensionless coupling constant~\cite{graphenereviews} (restoring $\hbar$ for a moment) 
$\alpha_{\rm ee} \equiv e^2/(\hbar v)$ which has a value $\approx 2.2$ in DLG and $  \approx 4.4$ in Bi$_2$Te$_3$ TIs if we use the respective Dirac velocities $v_{\rm G} \approx 10^{6}~{\rm m}/{\rm s}$ 
and $v_{\rm TI} \approx 5 \times 10^{5}~{\rm m}/{\rm s}$.

Several important many-body properties of the Hamiltonian ${\hat {\cal H}}$ are completely determined by the $2 \times 2$ 
symmetric matrix ${\bm \chi}(q,\omega)$ whose elements are the density-density linear-response functions
\begin{equation}\label{eq:LRT}
\chi_{\ell\ell'}(q,\omega) = \frac{1}{S} \langle \langle {\hat \rho}_{{\bm q}, \ell}; {\hat \rho}_{-{\bm q}, \ell}\rangle\rangle_\omega~,
\end{equation}
with $\langle\langle {\hat A},{\hat B}\rangle\rangle_\omega$ the usual Kubo product.~\cite{Giuliani_and_Vignale}  Within the RPA these functions satisfy the following matrix equation,
\begin{equation}\label{eq:matrix-form}
{\bm \chi}^{-1}(q,\omega) = {\bm \chi}^{-1}_0(q,\omega) - {\bm V}(q)~,
\end{equation}
where ${\bm \chi}_0(q,\omega)$ is a $2 \times 2$ diagonal matrix whose elements $\chi^{(0)}_\ell(q,\omega)$ are the well-known~\cite{hwang_prb_2007,barlas_prl_2007,wunsch_njp_2006} noninteracting (Lindhard) response functions of each layer at arbitrary doping $n_\ell$.  The off-diagonal (diagonal) elements of the matrix ${\bm V} =\{V_{\ell \ell'}\}_{\ell,\ell' = 1,2}$ represent inter-layer (intra-layer) Coulomb interactions. 

The bare intra- and inter-layer Coulomb interactions are influenced by the layered
 dielectric environment (see Fig.~\ref{fig:one}). A simple electrostatic calculation~\cite{profumo_prb_2010} implies that the Coulomb interaction in the $\ell =1$ (top) layer is given by
\begin{equation}\label{eq:v11}
V_{11}(q) = \frac{4\pi e^2}{q D(q)} [ (\epsilon_2 + \epsilon_3) e^{qd} + 
 (\epsilon_2 - \epsilon_3) e^{-qd}]~,
\end{equation}
where
\begin{equation}\label{eq:denominator}
D(q) =  [(\epsilon_1 + \epsilon_2) (\epsilon_2 + \epsilon_3) e^{qd} + 
(\epsilon_1 - \epsilon_2) (\epsilon_2 - \epsilon_3) e^{-qd} ]~.
\end{equation}
The Coulomb interaction in the bottom layer, $V_{22}(q)$, can be simply obtained from $V_{11}(q)$ 
by interchanging $\epsilon_3 \leftrightarrow \epsilon_1$. Finally, the inter-layer Coulomb interaction is given by
\begin{equation}\label{eq:v12} 
V_{12}(q) = V_{21}(q) = \frac{8\pi e^2}{q D(q)}~\epsilon_2~.
\end{equation}
Notice that in the ``uniform" $\epsilon_1 = \epsilon_2 = \epsilon_3 \equiv \epsilon$ limit we recover the familiar expressions 
$V_{11}(q) = V_{22}(q) \to 2\pi e^2/(\epsilon q)$ and $V_{12}(q) = V_{21}(q) \to V_{11}(q)\exp(-qd)$. 
Previous work on TI thin film and DLG collective modes has assumed this limit, which 
rarely applies experimentally.

\section{Collective modes}
\label{sect:collectivemodes}

The collective modes of the system described by the model Hamiltonian (\ref{eq:Hamiltonian}) can be 
determined by locating the poles of 
${\bm \chi}(q,\omega)$ in Eq.~(\ref{eq:matrix-form}).
A straightforward inversion of Eq.~(\ref{eq:matrix-form}) yields the following condition~\cite{dassarma_prb_1981,santoro_prb_1988}:
\ber\label{eq:zeroes}
\varepsilon(q,\omega) &=& [1-V_{11}(q) \chi^{(0)}_1(q,\omega)][1-V_{22}(q)\chi^{(0)}_2(q,\omega)] \nonumber\\
&-& V^2_{12}(q)\chi^{(0)}_1(q,\omega)\chi^{(0)}_2(q,\omega) =0~.
\eer
The collective modes occur above the intra-band particle-hole 
continuum where $\chi^{(0)}$ is real, positive, and a decreasing function of 
frequency.  Eq.~(\ref{eq:zeroes}) admits two solutions, a higher frequency 
solution~\cite{hwang_prb_2007,wunsch_njp_2006,polini_prb_2008} at $\omega_{\rm op}(q)$ which corresponds to in-phase oscillations of the 
densities in the two layers, and a lower frequency solution at 
$\omega_{\rm ac}(q)$ which corresponds to out-of-phase oscillations.

The plasmon collective modes of MD2DESs are of special interest because of the ease with which 
they may be altered by changing the carrier densities in either layer using gates.
We note in particular that the carrier densities in different layers can easily 
differ radically.  For this reason we present our results in terms of  
the total 2D carrier density $n = n_1 + n_2$, and the density polarization $\zeta = (n_2 - n_1)/n \in [-1,1]$: $\zeta = 1$ when the carrier density is non-zero only in the bottom layer 
($n_1 =0$), while $\zeta = 0$ when the two layers have identical carrier densities ($n_1 = n_2$). 

\subsection{Analytical results}
\label{sect:analyticalresults}

In this Section we report on
exact analytical expressions for the RPA optical and acoustic plasmon dispersions $\omega_{{\rm op}, {\rm ac}}(q)$
that are valid in the long-wavelength $q \to 0$ limit 
where $\omega_{\rm op}(q \to 0) \propto \sqrt{q}$ and $\omega_{\rm ac}(q \to 0) \propto q$.

We start by deriving an exact expression for the RPA long-wavelength acoustic-plasmon group velocity, 
\begin{equation}\label{eq:velocity}
c_{\rm s}= \lim_{q \to 0} \frac{\omega_{\rm ac}(q)}{q}~.
\end{equation}
Following Santoro and Giuliani~\cite{santoro_prb_1988}, we first introduce the power expansion
\be
\omega_{\rm ac}(q) = c_{\rm s} q + c_2 q^2 + c_3 q^3 + \dots
\ee
for the acoustic-plasmon dispersion relation, and then define a function
\be\label{eq:Ffunction}
F(q) = \varepsilon(q,c_{\rm s} q + c_2 q^2 + c_3 q^3 + \dots)~.
\ee
In the limit $q \to 0$ the function $F(q)$ has the following Laurent-Taylor expansion
\be
F(q) = f_{-1}~q^{-1} + f_0 + f_1~q + f_2~q^2 + \dots~,
\ee
where the coefficients $f_i$ can be extracted
from the analytical expression~\cite{barlas_prl_2007,hwang_prb_2007,wunsch_njp_2006} for the MD2DES 
Lindhard function $\chi^{(0)}_{\ell}(q,\omega)$. 
For Eq.~(\ref{eq:zeroes}) to be valid we have to require that the coefficients $f_i$ vanish identically. The coefficient $f_{-1}$ depends only on $c_{\rm s}$ and by equating its expression to zero we arrive after some tedious but straightforward algebra at the following equation for $x = c_{\rm s}/v$, the ratio between the  plasmon group velocity $c_{\rm s}$ 
and the Dirac velocity $v$:
\ber\label{eq:groupimplicit}
&& 2 g_{\rm s} g_{\rm v} \alpha_{\rm ee} {\bar d} (\zeta^2 - 1)[1+2x(\sqrt{x^2-1} - x)] \nonumber \\
& -& \sqrt{2}\epsilon_2 [1+x(\sqrt{x^2-1} - x)]f(\zeta) = 0~,
\eer
where $g_{\rm s}$ ($g_{\rm v}$) are real-spin (valley) degeneracy factors. In the case of DLG, $g_{\rm s} = g_{\rm v} = 2$, while in the case of thin-film TIs $g_{\rm s} = g_{\rm v} = 1$.  In Eq.~(\ref{eq:groupimplicit}) ${\bar d} = d k_{\rm F}$ is a dimensionless inter-layer distance calculated with $k_{\rm F}  \equiv \sqrt{4 \pi n/(g_{\rm s} g_{\rm v})}$ 
and $n = n_1 + n_2$, and 
\be\label{eq:fofzeta}
f(\zeta) = (1+\zeta)\sqrt{1-\zeta} + (1-\zeta)\sqrt{1+\zeta}~.
\ee 
Eq.~(\ref{eq:groupimplicit}) can be conveniently 
solved for $x$ by making the change of variables $x \mapsto \Gamma = \sqrt{x^2-1} - x$.
After some straigthforward algebra we find that 
\be\label{eq:soundvelocity}
\frac{c_{\rm s}}{v} = \frac{1 + \Lambda(\alpha_{\rm ee}{\bar d}/\epsilon_2, \zeta)}{[1 + 2\Lambda(\alpha_{\rm ee}{\bar d}/\epsilon_2, \zeta)]^{1/2}}
\ee
with
\be\label{eq:Lambdafunction}
\Lambda(\alpha_{\rm ee}{\bar d}/\epsilon_2, \zeta) = \frac{g_{\rm s} g_{\rm v}\sqrt{2}(1 - \zeta^2)}{f(\zeta)}\frac{\alpha_{\rm ee} {\bar d}}{\epsilon_2}~.
\ee 

Eqs.~(\ref{eq:soundvelocity})-(\ref{eq:Lambdafunction}) are the principle results of this Article. 
We see from this analytic expression that $c_{\rm s}$ is 
independent of $\epsilon_1$ and $\epsilon_3$ and depends only on the barrier material 
dielectric constant, which in the case of TI thin films is simply the TI bulk dielectric constant.
This behavior is a consequence of the out-of-phase character of this collective mode 
in which the double-layer total charge is locally constant but shifts dynamically between layers.  
Because TIs tend to have narrow gaps they tend to have large dielectric constants
($\epsilon_2 \sim 100$ in the case\cite{BiTkappa} of Bi$_2$Te$_3$).  Thin-film collective modes will therefore tend to have $c_{\rm s}/v$ values that are quite close to $1$ unless 
${\bar d}$ is very large.  (For large ${\bar d}$ the long-wavelength limit formula, which 
applies when both $qd$ and $q/k_{\rm F}$ are small, will have a limited range of applicability.)  

It follows from Eq.~(\ref{eq:soundvelocity}) that the ratio $c_{\rm s}/v$ is larger than unity for any value of the parameters 
$\alpha_{\rm ee}$, ${\bar d}$, $\zeta$, and $\epsilon_2$ so that the acoustic plasmon 
always lies outside of the MD2DES particle-hole continuum.
This implies than the acoustic plasmon is strictly speaking never 
Landau damped at small $q$.  
(A similar conclusion was reached previously~\cite{santoro_prb_1988} 
for the case of conventional 2D electron gases, but was limited to the 
case of identical density and hence identical Fermi velocity.)
 
For moderate values of $\bar{d}$, however, Eq.~(\ref{eq:soundvelocity}) 
predicts a TI thin film sound velocity so close to the top of the particle-hole continuum 
that it will likely be unobservable because of damping effects not 
captured by the RPA, and because of disorder, which is 
always present to some degree.
For the case of DLG, on the other hand, we expect that acoustic plasmon 
collective modes will be well defined.  This is particularly true in the case of 
DLG with a small number of layers of BN as barrier material.  When the 
BN barrier layer is very thin, the use of macroscopic dielectric parameters to characterize 
its screening properties is approximate; in that case measurement of the 
acoustic plasmon group velocity combined with 
Eqs.~(\ref{eq:soundvelocity})-(\ref{eq:Lambdafunction}) would allow the effective value of $\epsilon_2$ to be determined experimentally.

\begin{figure}
\centering
\includegraphics[width=1.00\linewidth]{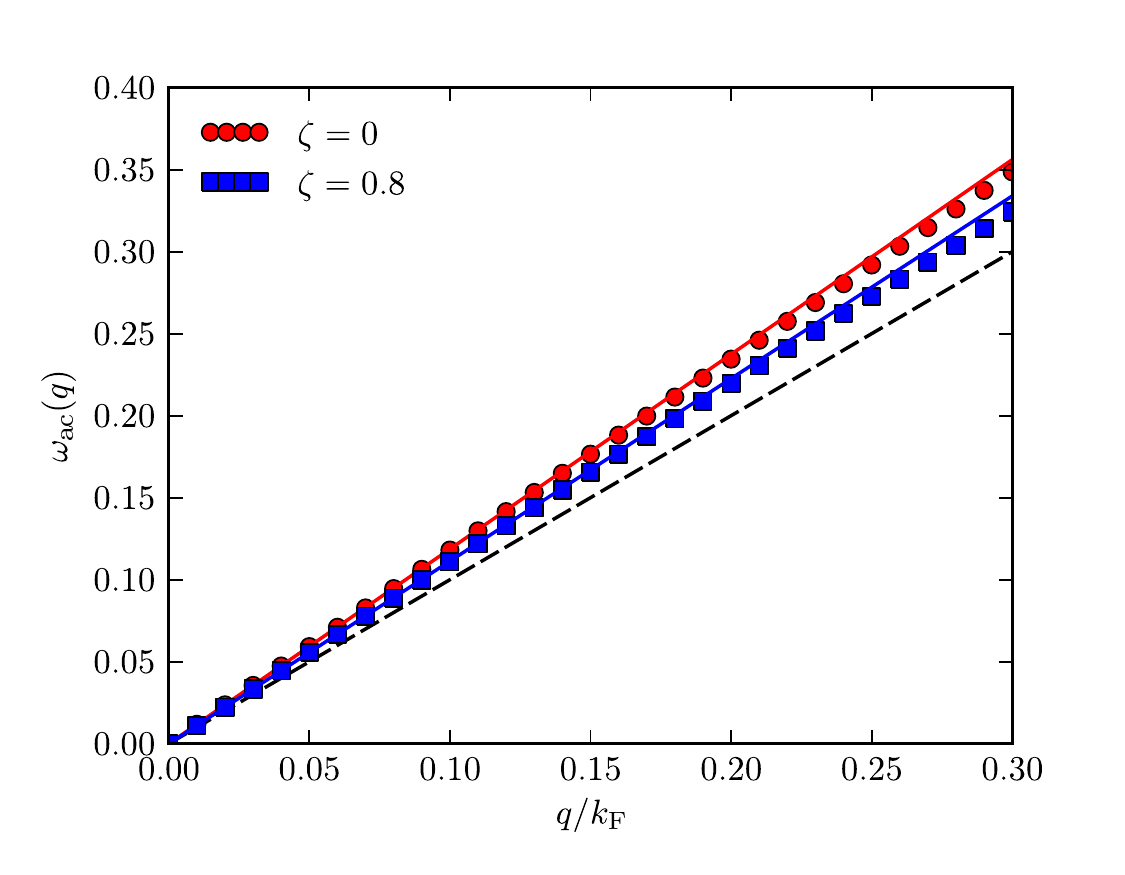}
\caption{(Color online) Long-wavelength acoustic plasmon dispersion of Coulomb-coupled massless-Dirac two-dimensional electron systems. 
The circles and squares are acoustic plasmon 
frequencies $\omega_{\rm ac}(q)$ (in units of $\varepsilon_{\rm F} = v k_{\rm F}$) as functions of $q/k_{\rm F}$  
calculated numerically from the solution of Eq.~(\ref{eq:zeroes}). Here 
$k_{\rm F}$ is the Fermi wave vector evaluated at the total density $n = n_1 + n_2$, {\it i.e.} $k_{\rm F} = \sqrt{4\pi (n_1 + n_2)/(g_{\rm s} g_{\rm v})}$. 
The parameters we have used to calculate the curve labeled 
by ``$\zeta =0$" are: $g_{\rm s} = g_{\rm v} =2$, $n_1 = n_2 = 5 \times 10^{12}~{\rm cm}^{-2}$, $\alpha_{\rm ee} = 2.2$, $d = 3.35~{\rm \AA}$, $\epsilon_1 = \epsilon_2 = 1$, and $\epsilon_3 = 3.9$.
These parameter values correspond to the case of two graphene layers on ${\rm SiO}_2$
that are decoupled by rotation. 
The data labeled by ``$\zeta=0.8$" have been calculated by setting $n_1 = 1 \times 10^{12}~{\rm cm}^{-2}$ 
and $n_2 = 9 \times 10^{12}~{\rm cm}^{-2}$ with the same values for the other parameters.
The solid lines plot $\omega = c_{\rm s}q$ for the $\zeta=0$ and $\zeta=0.8$ cases 
with the plasmon group velocity $c_{\rm s}$ calculated from the analytical result,
Eq.~(\ref{eq:soundvelocity}).  These numerical results confirm the validity of our analytic 
result for $c_{\rm s}$ and the importance of accounting for the delicate dependence of the 
long-wavelength Lindhard function on $\nu = \omega/(vq)$.  The dashed line plots the upper-bound of the intra-band electron-hole continuum,  $\omega = v q$. 
\label{fig:two}}
\end{figure}
 
We note that an analytic result for $c_{\rm s}$ was reported previously in 
Ref.~\onlinecite{hwang_prb_2009} [see their Eq.~(5b)] for the special case of 
DLG embedded in a uniform dielectric, {\it i.e.} for $\epsilon_1 = \epsilon_2 = \epsilon_3$. 
In our notation, their result reads
\be\label{eq:HDS}
\left.\frac{c_{\rm s}}{v}\right|_{\rm HDS} =  \left[ \frac{2\sqrt{2}\sqrt{1-\zeta^2}}{\sqrt{1-\zeta} +\sqrt{1+\zeta}}\frac{\alpha_{\rm ee} {\bar d}}{\epsilon_2}\right]^{1/2}~.
\ee
This equation is evidently different from Eq.~(\ref{eq:soundvelocity}) above. We believe that Eq.~(\ref{eq:soundvelocity}) is the correct RPA result for the acoustic-plasmon group velocity and that Eq.~(\ref{eq:HDS}) is incorrect.  The difference is due to the singular behavior of the Lindhard function $\chi^{(0)}_\ell(q,\omega)$ as 
a function of wave vector $q$ and frequency $\omega$ in the region in which both these quantities are small.
(See Sect.~4.4.3 of Ref.~\onlinecite{Giuliani_and_Vignale}.)  In particular,  the limit of $\chi^{(0)}_\ell(q,\omega)$ for $q \to 0$ and $\omega \to 0$ depends on the ratio $\nu = \omega/(vq)$, {\it i.e.} on the direction along which the origin of the 
$(q,\omega)$ plane is approached: different limits are obtained for different values of $\nu$. In an acoustic plasmon, the ratio $\nu$ approaches a constant as $q \to 0$ and thus the limit of $\chi^{(0)}_\ell(q,\omega)$ which matters is the one in which $q \to 0$ while 
the ratio $\omega/q$ is kept constant. This is the limit we have taken~\cite{santoro_prb_1988} in the derivation of Eq.~(\ref{eq:soundvelocity}) -- see Eq.~(\ref{eq:Ffunction}).   Eq.~(\ref{eq:HDS}) is obtained by incorrectly letting $q \to 0$ while $\omega$ is kept constant [see Eq.~(4) in Ref.~\onlinecite{hwang_prb_2009}]: in this limit $\nu$ {\it diverges} instead of going to a constant value. 

A careful comparison between our analytical prediction in Eq.~(\ref{eq:soundvelocity}) and the result obtained by the brute-force numerical solution of Eq.~(\ref{eq:zeroes}) is shown in Fig.~\ref{fig:two}. 
We clearly see that Eq.~(\ref{eq:soundvelocity}) compares very well with the full numerical result.

The analytical analysis of the long-wavelength optical plasmon mode is simpler since this 
mode satisfies $\omega_{\rm op}(q) \propto \sqrt{q}$ for $q \to 0$ and therefore occurs at $\nu  = \omega/(v q) \to \infty $.
We obtain an analytic result using the 
well-known high-frequency ($\omega \gg v q$ and $\omega \ll 2\varepsilon_{{\rm F}, \ell}$) 
dynamical limit of $\chi^{(0)}_\ell(q,\omega)$:
\be\label{eq:dynamical}
\lim_{q \to 0} \chi^{(0)}_\ell(q,\omega) = g_{\rm s} g_{\rm v}\frac{\varepsilon_{{\rm F}, \ell}}{4\pi} \frac{q^2}{\omega^2}~,
\ee
with $\varepsilon_{{\rm F}, \ell} = v k_{{\rm F}, \ell} = v \sqrt{4 \pi n_\ell/(g_{\rm s} g_{\rm v})}$. 
Using Eq.~(\ref{eq:dynamical}) in Eq.~(\ref{eq:zeroes}) we immediately find
\be\label{eq:opticalplasmon}
\omega^2_{\rm op}(q \to 0) = \frac{g_{\rm s} g_{\rm v}\alpha_{\rm ee}}{2 \bar{\epsilon}} v^2k_{\rm F}\left(\sqrt{\frac{1+\zeta}{2}} +\sqrt{\frac{1-\zeta}{2}}\right) q~,
\ee
with $\bar{\epsilon} = (\epsilon_1 + \epsilon_3)/2$.
Note that Eq.~(\ref{eq:opticalplasmon}) does not depend on the inter-layer distance 
or on the dielectric constant $\epsilon_2$, but only on the average $\bar{\epsilon}$ between top and bottom dielectric constants. Notice also that, in the limit $n_1 \to 0$ ($\zeta = 1$), Eq.~(\ref{eq:opticalplasmon}) reduces to the well-known plasmon frequency in a single-layer graphene sheet~\cite{polini_prb_2008,hwang_prb_2007,wunsch_njp_2006} with electron density $n_2$.
This expression applies for $qd \ll 1$, in which case the entire double-layer MD2DES acts in the 
optical plasmon mode like a single conducting layer at the interface between dielectric media 
characterized by constants $\epsilon_1$ and $\epsilon_3$.

\begin{figure}
\centering
\includegraphics[width=1.00\linewidth]{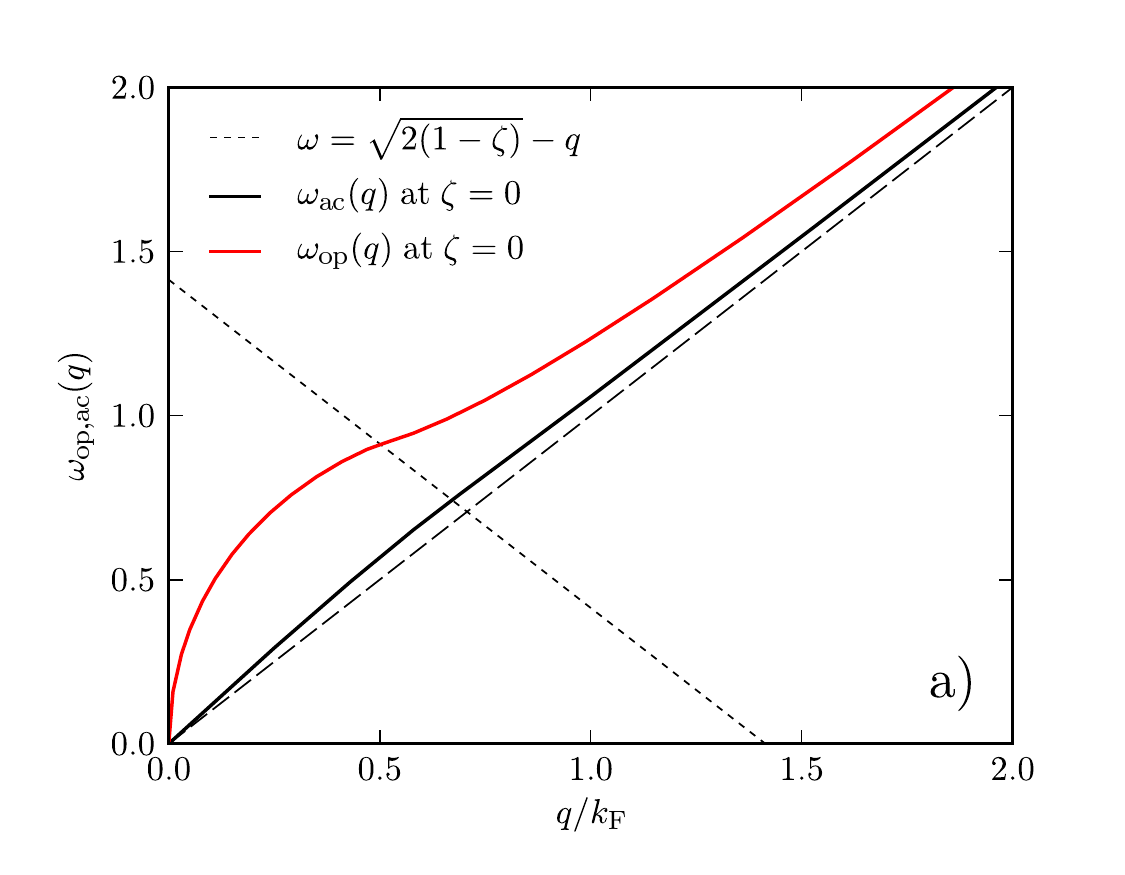}\\
\includegraphics[width=1.00\linewidth]{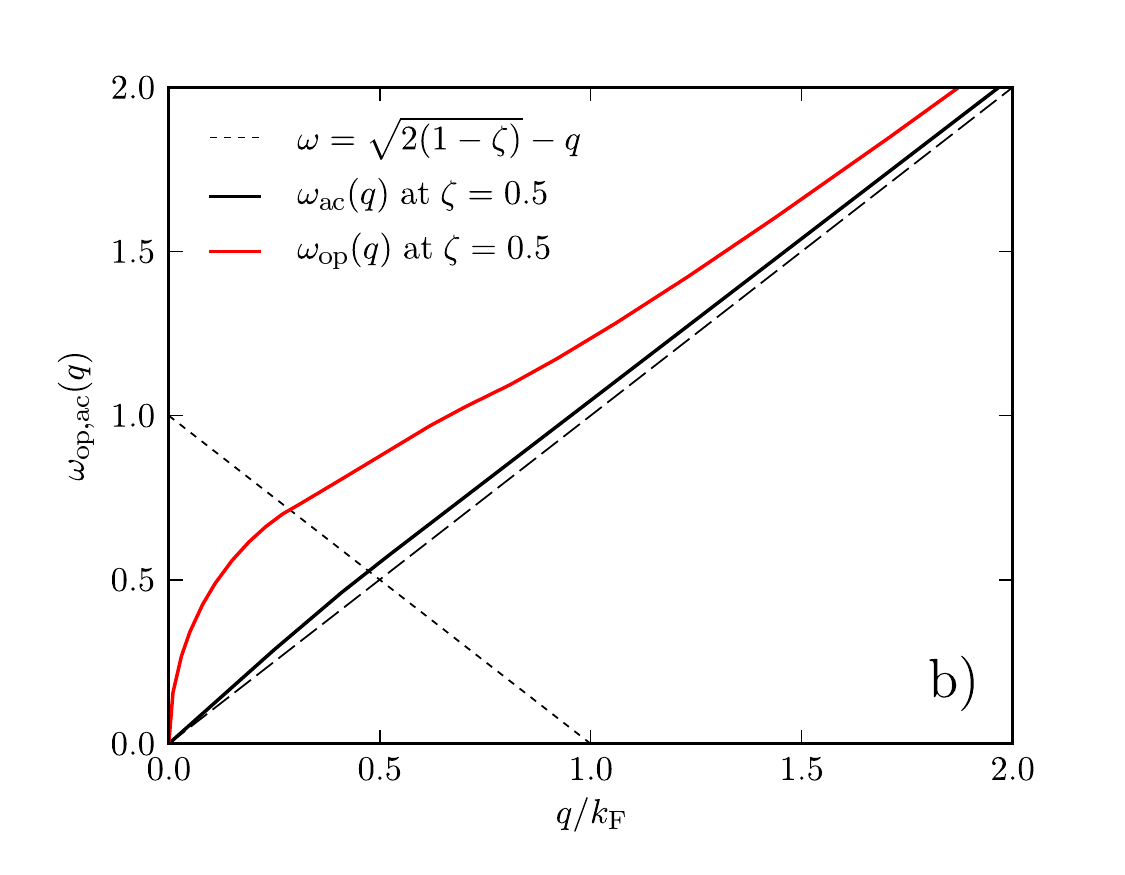}
\caption{(Color online) Panel a) Optical and acoustic plasmon dispersions (in units of the Fermi energy $\varepsilon_{\rm F} = v k_{\rm F}$) 
in a twisted double-layer graphene system on a ${\rm SiO}_2$ substrate as functions of wave vector $q$ [in units of $k_{\rm F} = \sqrt{4\pi (n_1 + n_2)/(g_{\rm s} g_{\rm v})}$]. The values of the parameters that we have used to produce the data in this figure are: $g_{\rm s} = g_{\rm v} =2$, $n_1 = n_2 = 5 \times 10^{12}~{\rm cm}^{-2}$ (corresponding to $n = 10^{13}~{\rm cm}^{-2}$ and $\zeta =0$), $\alpha_{\rm ee} = 2.2$, $d = 3.35~{\rm \AA}$, $\epsilon_1 = \epsilon_2 = 1$, and $\epsilon_3 = 3.9$. The intersections between the plasmon dispersions and the short-dashed line give the critical wave vector $q_{\rm c}$ at which Landau damping starts. Panel b) Same as in panel a) but for $\zeta =0.5$ [in producing the data shown in panel b) we have fixed the total density at the value used to produce the data in panel a), {\it i.e.} $n = 10^{13}~{\rm cm}^{-2}$]. 
\label{fig:three}}
\end{figure}

%
\subsection{Numerical results}
\label{sect:numericalresults}

In this Section we briefly report some representative numerical results for the optical and acoustic plasmon dispersion
relations obtained by solving Eq.~(\ref{eq:zeroes}), discussing first DLG and then TI thin films.  

In Fig.~\ref{fig:three} we illustrate the typical properties of DLG plasmon modes for the case
with the smallest MD2DES separation, two adjacent layers 
on a ${\rm SiO}_2$ substrate ($\epsilon_1 = \epsilon_2 = 1$ and $\epsilon_3 = 3.9$) that 
are weakly hybridized {\it e.g.} because of a twist between their orientations~\cite{twistedgraphene}. 
Fig.~\ref{fig:three}a) is for a symmetric system with the same electron concentration on the two layers ($\zeta = 0$), while Fig.~\ref{fig:three}b) refers to a system with a $50\%$ density imbalance. 
The characteristic behaviors $\omega_{\rm op}(q) \propto \sqrt{q}$ of the optical plasmon
and $\omega_{\rm ac}(q) \propto q$ of the acoustic plasmon are clearly visible.
The collective modes are not Landau damped when they appear in the gap between intra-band and inter-band particle-hole 
continua.  When the two layers have different densities, their particle-hole continua are different and the 
gap is smaller for the lower density layer. For adjacent but twisted DLG systems ${\bar d}$ is small even when the carrier density is large 
($\bar{d} \approx 0.2$ in Fig.~\ref{fig:three}).  It follows that $qd$ is small and the two MD2DESs are strongly coupled 
over the entire relevant frequency regime.  In this small ${\bar d}$ example the acoustic plasmon 
frequency is close to the particle-hole continuum because the capacitive energy associated with 
charge sloshing between the layers is proportional to the small layer separation.     

\begin{figure}
\centering
\includegraphics[width=1.00\linewidth]{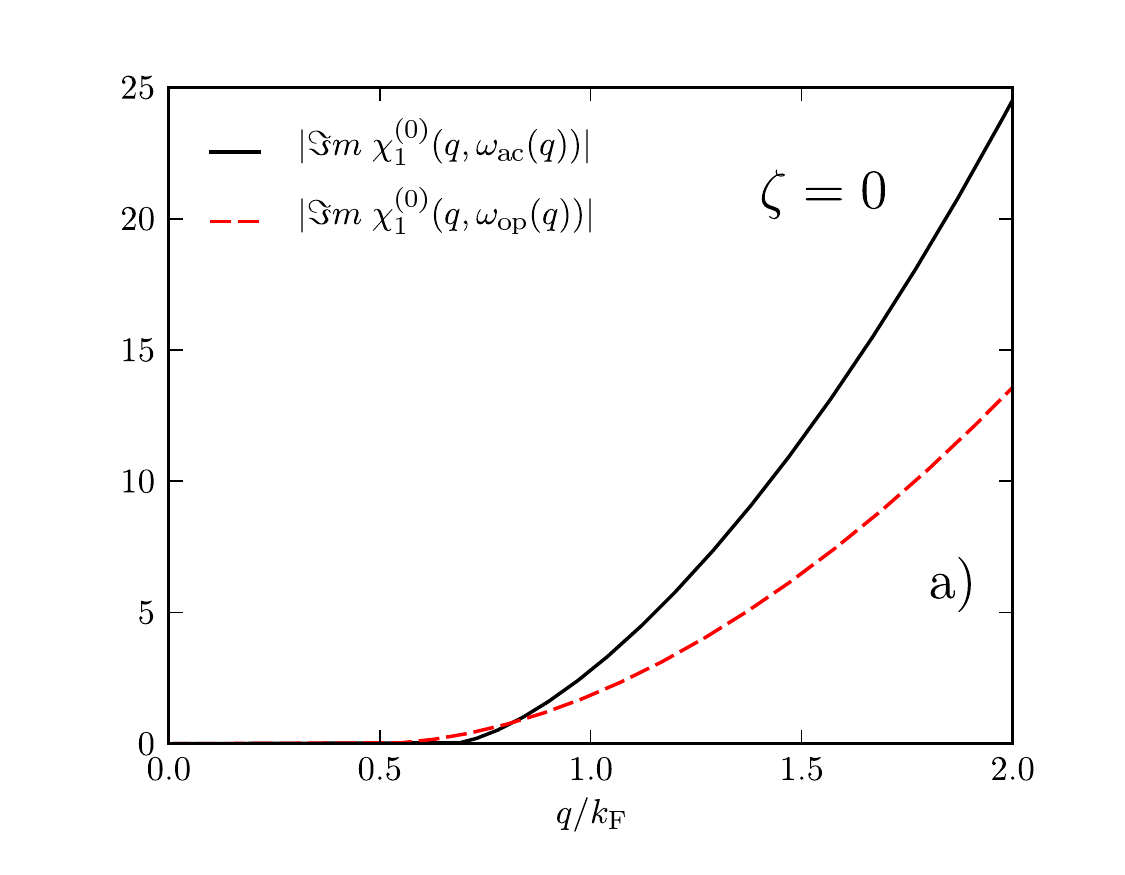}\\
\includegraphics[width=1.00\linewidth]{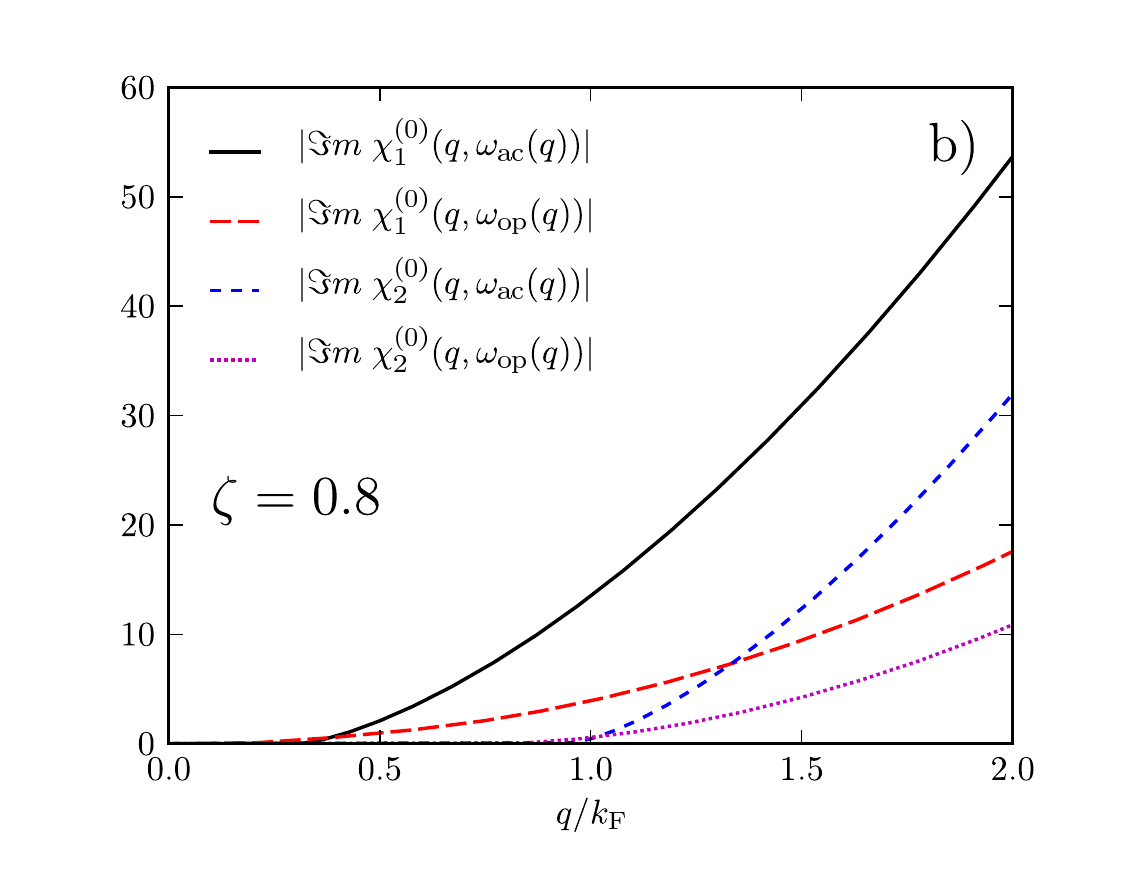}
\caption{(Color online) Landau damping of collective modes in double-layer graphene. 
Panel a) The absolute value of the imaginary part of the Lindhard function of the top layer, $|\Im m~\chi^{(0)}_1(q,\omega)|$, 
evaluated at the frequency $\omega = \omega_{\rm op}(q)$ 
[$\omega = \omega_{\rm ac}(q)$] of the optical [acoustic] plasmon. 
The data in this plot refer exactly to the parameters used in Fig.~\ref{fig:three}a). Note that, within RPA, $\Im m~\chi^{(0)}_1(q,\omega_{{\rm op}, {\rm ac}}(q))$ is identically zero for wave vectors $q$ 
up to a critical value $q^\star_{{\rm op}, {\rm ac}}$ at which $\omega_{{\rm op}, {\rm ac}}(q)$ hits the inter-band electron-hole continuum associated with the low-density layer. 
Since data in this panel correspond to $\zeta =0$, top-layer and bottom-layer Lindhard functions are identical. Panel b) Same as in panel a) but for $\zeta =0.8$ ($n_1 = 1 \times 10^{12}~{\rm cm}^{-2}$ 
and $n_2 = 9 \times 10^{12}~{\rm cm}^{-2}$). Note that the $q^\star_{{\rm op}, {\rm ac}}$ decreases with increasing $\zeta$ becoming zero in the limit $\zeta \to 1$. 
Since in this panel $\zeta \neq 0$ we have plotted both top-layer (low-density) and bottom-layer (high-density) Lindhard functions.\label{fig:four}}
\end{figure}

In Fig.~\ref{fig:four} we illustrate the strength of 
plasmon decay by emission of single electron-hole pairs (Landau damping).
Notice that Landau damping occurs when the curves $\omega_{{\rm op}, {\rm ac}}(q)$ in Fig.~\ref{fig:three} hit the inter-band electron-hole continuum of the layer with lower density (layer ``1" in our convention). The larger $\zeta$, the sooner this happens. In particular, in the limit in which layer ``1" is neutral ($\zeta =1$), Landau damping 
is present from vanishingly small wave vectors: damping of the optical plasmon excitation associated with electrons in the 
high-density layer starts at arbitrarily small wave vectors since decay
can easily occur {\it via} the emission of inter-band electron-hole pairs in the neutral layer. 
The many-body properties of two or more decoupled graphene layers can thus be strongly affected
by inter-layer Coulomb interactions, even by apparently innocuous
geometric features such as the presence of a nearly-neutral layer. 

\begin{figure}
\centering
\includegraphics[width=1.00\linewidth]{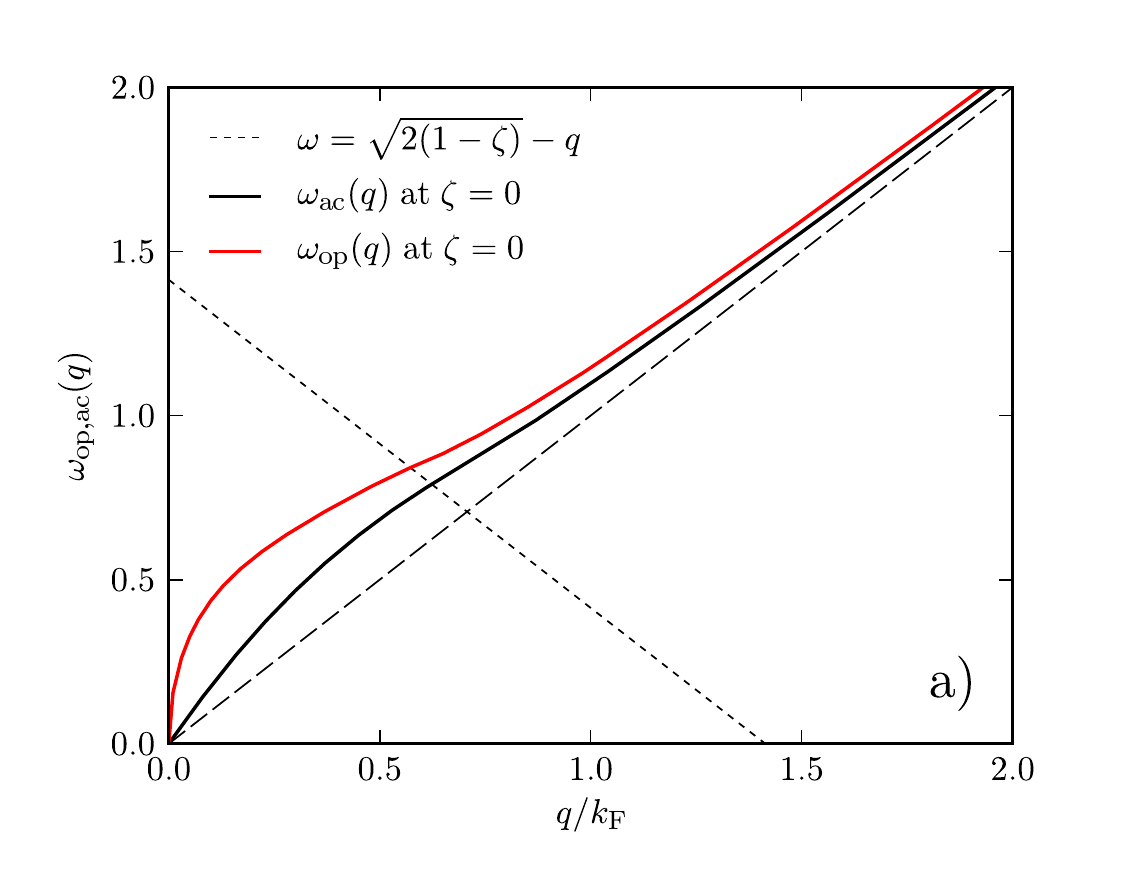}\\
\includegraphics[width=1.00\linewidth]{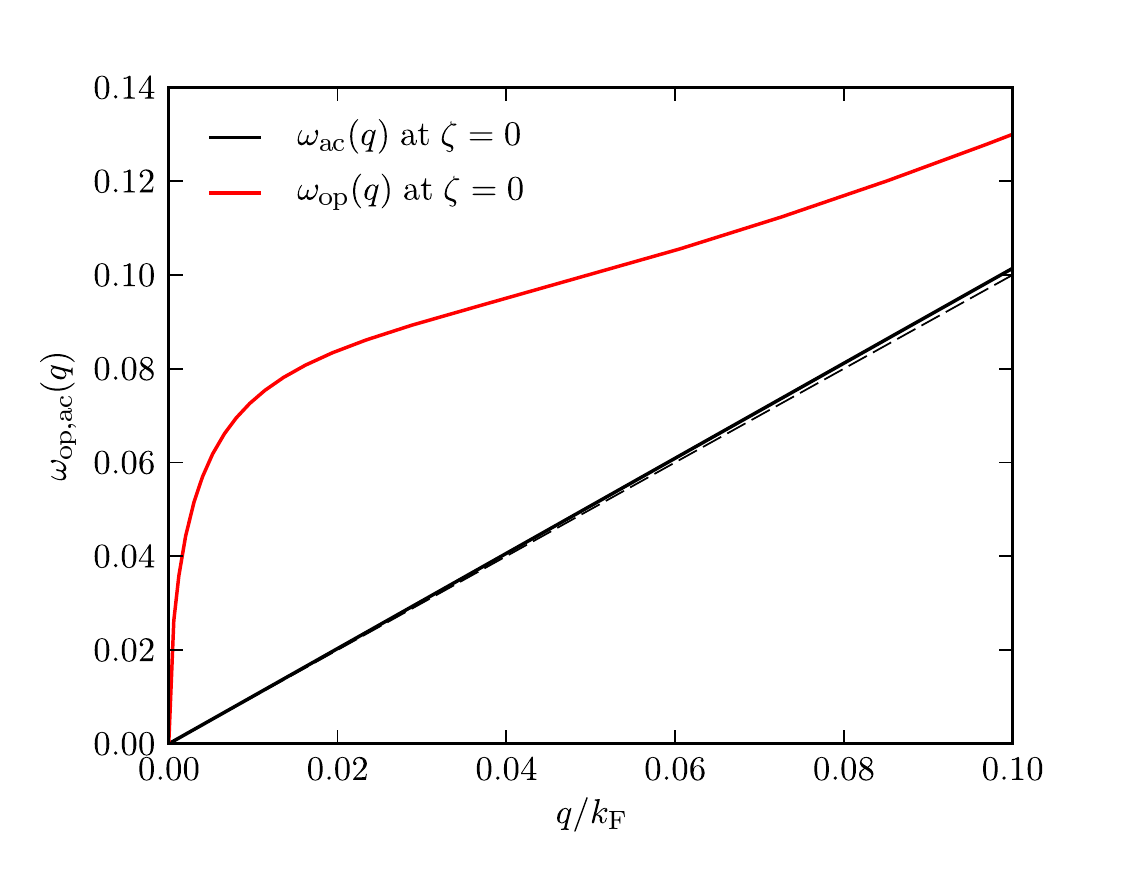}
\caption{(Color online)  Optical and acoustic plasmon dispersions (in units of the Fermi energy $\varepsilon_{\rm F} = v k_{\rm F}$) 
in a double-layer graphene system [panel a)] and a topological insulator thin-film [panel b)] as functions of wave vector $q$ [in units of $k_{\rm F} = \sqrt{4\pi (n_1 + n_2)/(g_{\rm s} g_{\rm v})}$]. 
The intersections between the plasmon dispersions and the short-dashed line give the critical wave vector $q_{\rm c}$ at which Landau damping starts.
Panel a) The values of the parameters that we have used to produce the data in this figure are: $g_{\rm s} = g_{\rm v} =2$, $n_1 = n_2 = 5 \times 10^{12}~{\rm cm}^{-2}$ (corresponding to $n = 10^{13}~{\rm cm}^{-2}$ and $\zeta =0$), $\alpha_{\rm ee} = 2.2$, $d = 6~{\rm nm}$, $\epsilon_1 = 1$, and $\epsilon_2 = \epsilon_3 = 4.0$. 
Panel b) The values of the parameters that we have used to produce the data in this figure are: $g_{\rm s} = g_{\rm v} =1$, $n_1 = n_2 = 5 \times 10^{12}~{\rm cm}^{-2}$, $\alpha_{\rm ee} = 4.4$, 
$d = 6~{\rm nm}$, $\epsilon_1 = 1$, $\epsilon_2 = 100$, and $\epsilon_3 = 4.0$. Note that due to the large value of the bulk TI dielectric constant, the acoustic plasmon is almost locked to the top of the intra-band electron-hole continuum.\label{fig:five}}
\end{figure}

In Fig.~\ref{fig:five} we compare optical and acoustic plasmon dispersions for DLG and TI thin-film systems.  
For the TI thin-film case we have chosen the following parameters: i) $\epsilon_1 = 1$, 
$\epsilon_2 =100$ (this roughly corresponds to the dielectric constant of ${\rm Bi}_2{\rm Te}_3$), and $\epsilon_3 = 4.0$; ii) a total electron density on the top and bottom surface states of $n  = 10^{13}~{\rm cm}^{-2}$; and iii) a thickness of the TI slab of $d = 6~{\rm nm}$, corresponding to a six quintile layer MBE-grown ${\rm Bi}_2{\rm Te}_3$ film (${\bar d} \approx 6.7$).  The DLG example has the 
same total density and layer separation  (${\bar d} \approx 3.4$; the difference in ${\bar d}$ in the two cases stems from the $g_{\rm s}$/$g_{\rm v}$ spin/valley degeneracy factors) and dielectric constants 
$\epsilon_1=1$ and $\epsilon_2=\epsilon_3=4.0$, corresponding to two graphene layers 
separated by approximately 15 BN layers and lying on a BN substrate.   
In both cases we see that a crossover occurs at intermediate values of $q$ between
strong (small $q$) and weak (large $q$) coupling of the two collective modes.
In the TI case the higher frequency optical plasmon mode deviates much more  
strongly from simple $\sqrt{q}$ behavior at this crossover because strong dielectric screening by the 
TI bulk suppresses the single-surface plasmon mode.  [Note however that the 
effective dielectric constant for this limit is $(\epsilon_2+\epsilon_{1,3})/2$ rather than 
$\epsilon_{2}$ as used in Ref.~\onlinecite{raghu_prl_2010}.]  The acoustic 
plasmon mode of the TI thin film case is, on the other hand, strongly
suppressed in the strong-coupling limit, as discusses earlier, and has 
a velocity much closer to the bare Dirac velocity than in the corresponding DLG case.

%
\section{Discussion and conclusions}
\label{sect:conclusions}

We have presented an analysis of the electronic
collective modes of systems composed of two 
unhybridized but Coulomb-coupled  
massless-Dirac two-dimensional electron systems (MD2DESs)
separated by a vertical distance $d$.
The primary example we have in mind is topological 
insulator (TI) thin films, which are always described at low energies 
by this type of model because topologically protected 
MD2DESs always appear on both top and bottom surfaces.
Also of interest are closely related systems, which we refer to 
as double-layer graphene (DLG) systems, containing two graphene 
layers that are weakly hybridized either because they are rotated relative 
to each other or because they are separated by a dielectric barrier layer.
Importantly, we allow for a general dielectric environment in which the 
material above the top MD2DES layer ($\epsilon_{1}$), between the two layers
($\epsilon_{2}$), and 
below the bottom MD2DES layer ($\epsilon_{3}$)
are all allowed to have different dielectric constants.
In the case of TI thin film $\epsilon_{2}$ is the bulk dielectric constant of the 
TI which is expected to have large values.   
The carrier collective modes of MD2DESs are expected to 
be most robust in the gap between intraband and interband particle-hole excitations.

The double-layer systems of interest quite generally have two collective modes
which in the limit of small $qd$ involve density-fluctuations in the two-layers that 
are strongly coupled, and in the limit of large $qd$ weakly coupled single-layer plasmons.
One key parameter which controls collective mode properties is the 
dimensionless product $k_{\rm F} d \equiv \bar{d}$.  Small values of $k_{\rm F}d$ imply that the 
layer separation is smaller than the typical distance between electrons within a layer and 
that collective modes at all values of $q$ up to $\sim k_{\rm F}$ are strongly coupled 
combinations of the two individual layer density-fluctuation contributions.  For large 
$k_{\rm F}d$ a crossover occurs for $q \in (0,k_{\rm F})$ between strongly and weakly coupled 
double-layer collective modes.  Both small and large values of $\bar{d}$ are achievable in samples 
where disorder plays an inessential role in both DLG and TI thin film cases.

Our study focuses on the long-wavelength limit in which both $qd$ and $q/k_{\rm F}=qd/\bar{d}$ are 
small.  We have derived analytic expressions for both frequencies of both the 
low-energy linearly dispersing acoustic plasmon mode $\omega_{\rm ac}(q)$ and for the 
high-energy optical plasmon mode $\omega_{\rm op}(q)$ which has $\sqrt{q}$ dispersion 
at long-wavelengths.  In this limit we find that $\omega_{\rm ac}(q) - vq \propto 1/\epsilon_{2}$ 
whereas $\omega_{\rm op}(q) \propto \sqrt{2/(\epsilon_{1}+\epsilon_{3})}$; {\it i.e.} the separation of the 
acoustic plasmon mode from the upper edge of the intra-band particle-hole continuum 
is very strongly suppressed by a large bulk TI dielectric constant, whereas the 
coupled double-layer plasmon mode is unaffected.  This double-layer optical plasmon behavior 
contrasts with that of a large $qd$ single-surface plasmon mode which has a frequency 
proportional to $\sqrt{2/(\epsilon_{2}+\epsilon_{1,3})}$. The long-wavelength limit of
$\omega_{\rm ac}(q)$ is sensitive not only to the energy associated with inter-layer charge sloshing 
but also to its microscopic kinetics as captured by the singular sensitivity 
of the MD2DES Lindhard function to $\omega/(v q)$.  By carefully accounting for this 
dependence we are able to correct a previous analytic expression in a way that 
is quantitatively particularly important in the TI thin film (large $\epsilon_{2}$) case.

Double-layer collective mode coupling plays an important role in MD2DES correlations when
$\bar{d}$ is small.  Even when $\bar{d}$ is large, strongly-coupled small $qd = q\bar{d}/k_{\rm F}$ modes 
will often be experimentally accessible and may play an important role in graphene multi-layer or 
TI based plasmonics.  The analytic results derived in this paper can be used to readily 
anticipate how these modes depend on system parameters.

From the more theoretical point of view, it will be intriguing to study physical properties of plasmons in Coulomb-coupled MD2DESs beyond the random phase approximation by employing {\it e.g.} many-body diagrammatic perturbation theory~\cite{abedinpour_prb_2011}.

\acknowledgments

Work in Pisa was supported by the Italian Ministry of Education, University, and Research (MIUR) through the program ``FIRB - Futuro in Ricerca 2010" (project title ``PLASMOGRAPH: plasmons and terahertz devices in graphene"). A.H.M. was supported by Welch Foundation Grant No. TBF1473, DOE Division of Materials Sciences and Engineering Grant No. DEFG03-02ER45958, 
and by the NRI SWAN program. M.P. acknowledges the kind hospitality of the IPM (Tehran, Iran) during the final stages of preparation of this work. 

While this manuscript was being finalized for publication, we became aware of a study of optical and acoustic plasmons in double-layer graphene~\cite{stauber_arXiv_2011}. The authors of this work present extensive numerical results for the ``uniform medium" limit ($\epsilon_1 = \epsilon_2 = \epsilon_3$) and discuss the relation between (longitudinal and transverse) plasmons and near-field amplification.

\end{document}